\documentclass[12pt]{article}
\usepackage{epsfig,psfrag,color}

\begin{document}

\begin{titlepage}

\begin{flushright}
{\small
CERN-PH-TH/2009-021\\
LMU-ASC~07/09\\
MZ-TH/09-04\\
PITHA~09/06\\
SHEP~0904\\
SFB/CPP-09-13\\
25 February 2009
%Draft \today
% arXiv:0902.nnnn [hep-ph]
}
\end{flushright}

\vspace{0.5cm}
\begin{center}
\Large\bf 
Penguins with Charm and Quark-Hadron Duality
\end{center}

\vspace{3mm}
\begin{center}
{\sc M. Beneke\,$^{1,2}$, G. Buchalla\,$^3$, M. Neubert\,$^4$, 
and C.T. Sachrajda\,$^5$}
\end{center}

\vspace{1mm}
\begin{center}
{\em 
$^1$\,Institut f\"ur Theoretische Physik E\\ 
RWTH Aachen University, D--52056 Aachen, Germany\\[3mm]
{$^2$\,CERN Theory Department\\
CH--1211 Gen\`{e}ve, Switzerland}\\[3mm]
$^3$\,Ludwig Maximilians-Universit\"at M\"unchen, 
Fakult\"at f\"ur Physik\\
Arnold Sommerfeld Center for Theoretical Physics, 
D--80333 M\"unchen, Germany\\[3mm]
$^4$\,Institut f\"ur Physik (THEP), Johannes Gutenberg-Universit\"at\\
D--55099 Mainz, Germany\\[3mm]
$^5$\,School of Physics and Astronomy, University of Southampton\\
Southampton SO17 1BJ, U.K.}
\end{center}

\vspace{1mm}
\begin{abstract}\noindent
The integrated branching fraction of the process $B\to X_s\,l^+l^-$ is dominated by resonance background from narrow charmonium states, such as $B\to X_s\psi\to X_s\,l^+l^-$, which exceeds the non-resonant charm-loop contribution by two orders of magnitude. The origin of this fact is discussed in view of the general expectation of quark-hadron duality. The situation in $B\to X_s\,l^+l^-$ is contrasted with charm-penguin amplitudes in two-body hadronic $B$ decays of the type $B\to\pi\pi$, for which it is demonstrated that resonance effects and the potentially non-perturbative $c\bar c$ threshold region do not invalidate the standard picture of QCD factorization. This holds irrespective of whether the charm quark is treated as a light or a heavy quark.
\end{abstract}

\end{titlepage}

\section{Introduction}

It is a well-known fact that the resonant transition $B\to X_s\psi\to X_s\,l^+l^-$ exceeds the non-resonant short-distance process $B\to X_s\,l^+l^-$ by about two orders of magnitude, even when considering the rate fully integrated over the dilepton invariant mass. Since the process is inclusive over the hadronic final states, this is sometimes interpreted as a particularly striking failure of global parton-hadron duality, which says that the sum over the hadronic final states, including resonances, should be well approximated  by a quark-level calculation \cite{Poggio:1975af}. The origin of the problem is apparently rooted in the fact that the process $B\to X_s\,l^+l^-$ contains a penguin contribution with a charm-quark loop, which has only a small effect on the partonic calculation but leads to a large resonant  enhancement in the real world. Charm-quark loops are relevant in other contexts as well, such as in the calculation of the charm production cross section in $e^+e^-$ annihilation from the imaginary part of a current correlator, and in the calculation of penguin amplitudes in non-leptonic $B$ decays in the QCD factorization framework \cite{BBNS1,BBNS2,BBNS3}. Should one expect large violations of global parton-hadron duality due to charmonium resonances in these cases as well? The purpose of this note is to explain the origin of the resonance dominance in $B\to X_s\,l^+l^-$ decays. We then investigate the differences between this process, where duality should not be expected to hold, and non-leptonic $B$ decays, where we demonstrate that it holds in the heavy-quark limit.

The contributions of charmonium states to the different observables studied in this note are described in terms of resonance contributions to a correlation function $\Pi(q^2)$. The question of when large deviations from quark-hadron duality arise is tightly linked to how this function enters the formulae for the various decay rates. After a brief review of some basic facts about $B\to X_s\,l^+ l^-$ decays, we study the correlator of two charm-quark currents using the operator product expansion (OPE) as well as a hadronic picture in terms of Coulomb resonance states. We then construct a toy model, which elucidates important features of the correlator and their relation to violations of quark-hadron duality. Returning to the case of $B\to X_s\,l^+ l^-$ decays, we derive numerical estimates explaining the observed large violation of duality in a semi-quantitative way. Finally, we consider the question of duality violations in charm-loop penguin contributions to exclusive hadronic decays such as $B\to\pi\pi$. Our discussion in this note focuses on the inclusive leptonic process $B\to X_s\,l^+l^-$, for which duality violations in the charm resonance region are particularly pronounced and experimentally well studied. We emphasize that an analogous discussion holds for the related radiative process $B\to X_s\gamma$, for which the resonant transition $B\to X_s\psi$ followed by $\psi\to X\gamma$ also exceeds the total inclusive short-distance contribution.

\section{\boldmath Brief review of $B\to X_s\,l^+l^-$ decays}

\begin{figure}[t]
\begin{center}
\resizebox{7.5cm}{!}{\includegraphics{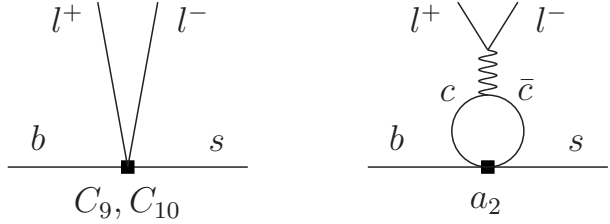}}
\caption{\label{fig:bslldiags}
Short-distance (left) and charm-penguin contributions (right) to $b\to s\,l^+l^-$.}
\end{center}
\end{figure}

The total inclusive branching fraction for the process $B\to X_s\,l^+l^-$ may be written as~\cite{BBL}
\begin{equation}\label{bsee}
   \mbox{B}(B\to X_s\,l^+l^-)_{\rm SD}
   = \tau_B\,\frac{G^2_F m^5_b}{192\pi^3}\,\frac{\alpha^2}{4\pi^2}\,
   |V_{ts}|^2\,\frac{\langle |C_9|^2\rangle + |C_{10}|^2}{2} 
   \approx 5.3\cdot 10^{-6} ,
\end{equation}
where we have neglected the contribution from the magnetic dipole operator $Q_{7\gamma}$, which is immaterial to our discussion. The subscript ``SD'' indicates that this expression is the short-distance contribution to the branching fraction, computed using partonic matrix elements of the effective weak Hamiltonian. The quantity $\langle |C_9|^2\rangle$ is understood to be the effective coefficient $|C^{\rm eff}_9|^2$, which includes the penguin-type matrix elements, averaged over the dilepton-mass squared:
\begin{equation}\label{c9av}
   \langle |C_9|^2\rangle 
   = 2\int_0^1\!ds\,(1-s)^2\,(1+2s)\,
   \left| C_9^{\rm NDR} + 3a_2\,h(z,s) \right|^2 ,
\end{equation}
where $h(z,s)$ with $z=m_c/m_b$ and $s=q^2/m^2_b$ is the charm-penguin loop function, and $a_2=C_2+C_1/3\approx 0.12$ \cite{BBL}.\footnote{Our coefficients $C_{1,2}$ are called $C_{2,1}$ in this reference.} 
The leading contributions to the decay rate are a large logarithmic term $\sim\ln(M_W/m_b)$ in $C_9$ from the ultraviolet contribution to the charm-penguin amplitude (high-energy scales between $m_b$ and $M_W$) and the top-quark contributions to $C_9$ and $C_{10}$. Since $C_9^{\rm NDR}\approx -C_{10}\approx 4$ and $h(z,s)$ is of order 1, we conclude that the low-energy contribution (scales at or below $m_b$) of the charm penguin is subleading. Numerically, $(\langle |C_9|^2\rangle+|C_{10}|^2)/2\approx 16$ in (\ref{bsee}). We also use $\alpha=\alpha(m_b)=1/132$ and $m_b=4.8$\,GeV. Feynman diagrams illustrating the short-distance and charm-loop contributions are shown in Figure~\ref{fig:bslldiags}.

On the other hand, the branching fraction for $B\to X_s\psi$ reads (with $r=M^2_\psi/m^2_b$)
\begin{equation}\label{bspsi}
   \mbox{B}(B\to X_s\psi) 
   = \tau_B\,\frac{G^2_F m^3_b f^2_\psi}{16\pi}\,
   |V_{cb} V_{cs}|^2\,a^2_2\,(1-r)^2\,(1+2r) 
   \approx 1.7\cdot 10^{-3} .
\end{equation}
This falls short of the experimental value $\mbox{B}(B\to X_s\psi)=(7.8\pm 0.4)\cdot 10^{-3}$ \cite{Amsler:2008zzb}, since the inclusive $\psi$ production rate is probably dominated by colour-octet production \cite{Ko:1995iv,Beneke:1998ks}. Phenomenologically, the experimental rate may be accounted for by introducing a factor $\kappa=2.16$ and replacing
\begin{equation}\label{kappaa2}
   a_2\to\kappa a_2 \,.
\end{equation}
Multiplying the experimental value with the leptonic branching fraction $\mbox{B}(\psi\to l^+l^-)=0.06$, we obtain the resonance contribution $\mbox{B}(B\to X_s\psi\to X_s\,l^+l^-)\approx 4.7\cdot 10^{-4}$ and thus 
\begin{equation}\label{bspsiee}
   R_\psi\equiv 
   \frac{\mbox{B}(B\to X_s\psi\to X_s\,l^+l^-)}%
        {\mbox{B}(B\to X_s\,l^+l^-)_{\rm SD}}
   \approx 90 \,.
\end{equation}
This estimate is only slightly modified by contributions from the higher charmonium resonances. Since the denominator includes the partonic charm-loop contribution, we conclude that the quark-level calculation of the {\em total\/} branching fraction yields a result that is far smaller than the contributions from the charmonium resonances alone, indicating a gross failure of global quark-hadron duality.

\begin{figure}[t]
\begin{center}
\psfrag{x}[t]{$s$}
\psfrag{y}[b]{$d\mbox{B}(B\to X_s\,l^+l^-)/ds~~[10^{-5}]$}
\resizebox{10cm}{!}{\includegraphics{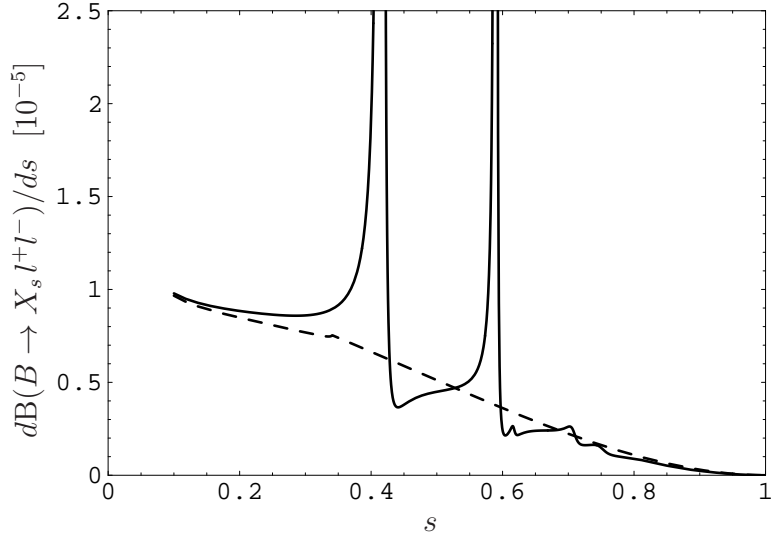}}
\caption{\label{fig:bsllplot}
Differential $B\to X_s\,l^+l^-$ branching fraction as a function of $s=q^2/m^2_b\equiv m_{l^+l^-}^2/m_b^2$, including the effect of charm resonances in the KS method (solid line). A phenomenological factor of $\kappa=2.16$ is used for the $\psi$ resonance amplitude. For comparison, the dashed curve shows the same quantity obtained within a purely partonic calculation. The short-distance contributions are computed at next-to-leading logarithmic order.}
\end{center}
\end{figure}

In order to include the resonance contributions in a phenomenological description of the global features of the $B\to X_s\,l^+l^-$ spectrum, Kr\"uger and Sehgal (KS) have suggested to relate the charm-penguin loop to the hadronic $c\bar c$ vacuum polarization function $\Pi(q^2)$ using a dispersion relation \cite{Kruger:1996cv}. This assumes that QCD interactions between the charm loop and the remaining quarks in the process can be neglected. On the other hand, the KS model offers the possibility to take resonances and other hadronic contributions related to the charm loop into account. Employing the dispersion relation gives the correct scheme dependence of the loop function and avoids a double counting of partonic and hadronic degrees of freedom. The ansatz of \cite{Kruger:1996cv} amounts to the replacement of the penguin function $h(z,s)$ by
\begin{equation}\label{hks}
   h_{KS} = - \frac{8}{9} \ln\frac{m_c}{\mu} - \frac{4}{9}
   + \frac{16\pi^2}{9} \left[ \Pi(q^2) - \Pi(0) \right] .
\end{equation} 
The vacuum polarization function $\Pi(q^2)$ will be defined in (\ref{pimunu}) below. Its imaginary part can be obtained from experimental data and related to $\Pi(q^2)$ through the dispersion relation (\ref{disprel}). An explicit parametrization with six Breit-Wigner resonances and a continuum contribution can be found in \cite{Kruger:1996cv}. To account for the experimental $B\to X_s\psi$ rate discussed after (\ref{bspsi}), KS include the phenomenological enhancement factor $\kappa$ in (\ref{kappaa2}), which increases the $\psi$ resonance term. The global $q^2$ spectrum for $B\to X_s\,l^+l^-$ decays with resonances included using the KS approach is shown in Figure~\ref{fig:bsllplot}.

From the preceding discussion it is clear that important information on the hadronic dynamics of $B\to X_s\,l^+l^-$ is contained in the vacuum polarization function, which we briefly review in the following section.

\section{Vector-current correlator}

\subsection{Parton-hadron duality}

In discussing the validity of a perturbative quark-level computation of the penguin amplitude, the concept of quark-hadron duality is crucial. We will therefore recall basic aspects of duality for the correlator of two vector currents. This is a particularly simple situation, but it is also relevant for the penguin amplitude in the approximation where further gluon attachments to the charm loop are neglected.

The correlator is defined by
\begin{equation}\label{pimunu}
   \Pi_{\mu\nu}
   = i\int\!d^4x\,e^{iq\cdot x}\,
   \langle 0|T\,j_\mu(x) j_\nu(0)|0\rangle
   \equiv (q_\mu q_\nu - q^2 g_{\mu\nu})\,\Pi(q^2) \,,
\end{equation}
where $j_\mu=\bar c\gamma_\mu c$ for the charm quark, and $c\to u,d,s$ for the case of light quarks. The scalar function $\Pi(q^2)$ obeys the dispersion relation
\begin{equation}\label{disprel}
   \Pi(q^2) - \Pi(0)
   = \frac{q^2}{\pi} \int_0^\infty\!\frac{dt}{t}\,
   \frac{ \mbox{Im}\,\Pi(t)}{t-q^2-i\epsilon} \,,
\end{equation}
which is defined in the Euclidean domain $q^2<0$ and may be analytically continued to the time-like region $q^2>0$. The expression on the left-hand side of (\ref{disprel}) can be computed in QCD by means of an OPE as a power series in $1/q^2$, with the coefficient of each term given as an expansion in the strong coupling $\alpha_s$. We shall denote this quark-level (OPE) result for $\Pi(q^2)$ as $\Pi_q(q^2)$. To lowest order in the OPE, and omitting ${\cal O}(\alpha_s)$ corrections, the correlator is given by
\begin{equation}\label{piquark}
   \Pi_q(q^2) - \Pi_q(0) 
   = - \frac{N}{2\pi^2} \int_0^1\!du\,u(1-u)\,
   \ln\frac{m^2-u(1-u)q^2-i\epsilon}{m^2}
   \approx -\frac{N}{12\pi^2}\,\ln\frac{-q^2-i\epsilon}{m^2} \,,
\end{equation}
with $m$ the quark mass and $N=3$ the number of colours. The second expression in (\ref{piquark}) is the asymptotic behaviour for large $q^2$. The imaginary part is non-zero only if $q^2>4m^2$ and reads
\begin{equation}\label{impiq}
    \mbox{Im}\,\Pi_q(q^2)
   = \frac{N}{12\pi}\,\sqrt{1-\frac{4m^2}{q^2}}
   \left( 1 + \frac{2m^2}{q^2} \right) .
\end{equation}
On the other hand, eq.~(\ref{disprel}) is valid for the exact, hadronic correlator $\Pi_h(q^2)$. The imaginary part of $\Pi_h$ is related to the observable $R$ ratio through\footnote{The relation is approximate, since $c\bar c$ pairs may annihilate or be produced from gluon splitting.}
\begin{equation}\label{rcimpih}
   R_c \equiv
   \frac{\sigma(e^+e^-\to\mbox{hadrons}\,(c\bar c))}%
        {\sigma(e^+e^-\to\mu^+\mu^-)}
   = 12\pi e_c^2\,\,\mbox{Im}\,\Pi_h \,,
\end{equation}
where $e_c=2/3$ is the charm-quark charge. Using (\ref{disprel}), $\Pi_h(q^2)$ may then, in principle, be obtained from experimental data.

To any finite order in $\alpha_s$ and $1/q^2$, the OPE quantity $\Pi_q(q^2)$ is expected to approximate the true function $\Pi_h(q^2)$ in the sense of an asymptotic expansion. For $q^2<0$, with sufficiently large $|q^2|$, the difference between $\Pi_q$ and $\Pi_h$ should then be systematically reduced with the inclusion of more terms in the OPE, up to a residual uncertainty, which is expected to be exponentially suppressed as $\exp(-c\,|q^2|/\Lambda_{\rm QCD}^2)$ with an ${\cal O}(1)$ constant $c$. For $q^2>0$ the deviation of the smooth function $\Pi_q(q^2)$ from $\Pi_h(q^2)$ might also involve oscillatory terms in $q^2$. The approximation of $\Pi_h$ by $\Pi_q$ will then in general not work point by point, but only if $q^2>0$ is large enough or $\mbox{Im}\,\Pi$ is averaged over a sufficiently wide range in $q^2$ \cite{Poggio:1975af}. The correspondence between $\Pi_q$ and $\Pi_h$ point by point in $q^2$ is referred to as local quark-hadron duality, the correspondence of averages as global quark-hadron duality. An illuminating discussion of these concepts can be found in \cite{Chibisov:1996wf,BSZ}. 

\subsection{Vector-current correlator in the Coulombic limit}

The concept of duality is of particular importance for light-quark correlators, where the resonance region is always non-perturbative. For massive quarks, we may also consider the opposite limit, where not only $2 m$, but also the binding energy of the quarkonium resonances is much larger than the strong-interaction scale. This limit is referred to as ``Coulombic'', since the dominant binding force is then the colour-Coulomb force. Although charmonium is not a Coulomb bound state, it is instructive to discuss the Coulombic limit of the charm-quark vector-current correlator in the present context. Some of the results of this section will also be used in the numerical estimates below.

Near $q^2\approx 4m^2_c$ the spectral function $\mbox{Im}\,\Pi_h(q^2)$ receives prominent contributions from a series of $c\bar c$ resonances (and other hadronic states with open charm), which cannot be captured by a fixed-order perturbative expansion. The exchange of Coulomb gluons gives rise to corrections of order $\alpha_s/v$, where $v$ is the velocity of the heavy quarks \cite{Sommerfeld}. In the threshold region $v\sim\alpha_s$ the entire ladder of Coulomb gluons needs to be resummed in order to account for the detailed behaviour of $\Pi(q^2)$. When this is done, one finds for the correlator after $\overline{\rm MS}$ subtractions \cite{Beneke:1999zr}
\begin{equation}\label{picoul}
   \Pi_C(q^2)
   = - \frac{\alpha_s C_F N}{8\pi} \left[ \frac{1}{2\lambda}
   + \frac{1}{2} \ln\frac{-4m_c E}{\mu^2} - \frac{1}{2}
   + \gamma_E + \psi(1-\lambda) \right] ,
\end{equation}
with
\begin{equation}\label{lambdae}
   \lambda = \frac{\alpha_s C_F}{2}\sqrt{\frac{m_c}{-E}} \,,
    \qquad
   E =\sqrt{q^2} - 2m_c +i\epsilon \,.
\end{equation}
Here $\psi(z)=d\ln\Gamma(z)/dz$ denotes the digamma function, and $C_F=(N^2-1)/(2N)$. Expression (\ref{picoul}), valid for $q^2\approx 4m_c^2$, contains poles for positive integer values $\lambda=n$, which correspond to the charmonium binding energies
\begin{equation}\label{ecoul}
   E_n = - \frac{m_c(C_F\alpha_s)^2}{4 n^2}
\end{equation}
in the colour-Coulomb potential. The residues of these poles determine the decay constants $f_n$ of the S-wave bound states at energy $E_n$. The pole part of (\ref{picoul}) is thus equivalent to
\begin{equation}\label{pipole}
   \Pi_{C,\,\rm pole}(q^2)
   = - \sum^\infty_{n=1} \frac{f^2_n}{q^2-M^2_n+i\epsilon} \,,
\end{equation}
where $M_n=2 m_c+E_n$ and $f^2_n=f^2_\psi/n^3$. The decay constant $f_\psi$ of the ground state is given in terms of the wave function at the origin by
\begin{equation}\label{fpsipsi}
   f^2_\psi = \frac{4N\,|\psi(0)|^2}{M_\psi}
    = \frac{16}{9\pi}\,m^2_c\,\alpha^3_s \,,
    \qquad
   |\psi(0)|^2 = \frac{1}{\pi} 
    \left( \frac{m_c}{2}\,C_F\alpha_s \right)^3 .
\end{equation}

The validity of local vs.\ global parton-hadron duality in the Coulombic limit is related to the issue of perturbative resummation. When sufficiently wide and smooth averages of $\mbox{Im}\,\Pi$ or $\Pi$ over the resonance region are taken, global duality holds in the sense that these quantities are well approximated by fixed-order perturbative and OPE computations without any need to sum the Coulomb ladders. In addition, in the Coulombic limit one may expect that local duality holds as well, since the physics of the threshold region is perturbative. It is evident that a local correspondence of $\mbox{Im}\,\Pi_h$ and $\mbox{Im}\,\Pi_q$ requires the perturbative resummation of Coulomb gluons discussed above. However, even in the ultra-heavy-quark limit there is a non-perturbative contribution from the strong-interaction scale to the charmonium binding energy \cite{Voloshin}. Thus, for local duality to hold this contribution must be small compared to the width of the resonance.

The widths of the resonances are not present in (\ref{picoul}) and (\ref{pipole}), since in the Coulombic limit they are of higher order in small coupling constants. A further resummation is required that replaces $M_n^2\to M_n^2-i M_n \Gamma_n$ in the denominator of (\ref{pipole}). Concentrating on $n=1$, the $\psi$ resonance can decay via the strong interaction into a three-gluon final state or into a fermion pair ($e$, $\mu$, $u$, $d$, $s$) through a virtual photon. The total decay rate of the ground state is then given by
\begin{equation}\label{gammapsi}
   \Gamma_\psi = \Gamma(\psi\to 3g) + 4\Gamma(\psi\to e^+e^-) \,,
\end{equation}
where
\begin{equation}\label{psiwidths}
   \Gamma(\psi\to 3g)
   = \frac{40(\pi^2-9)\,\alpha^3_s}{243}\,\frac{f^2_\psi}{M_\psi} \,,
    \qquad
   \Gamma(\psi\to e^+e^-)
   = \frac{16\pi\alpha^2 f^2_\psi}{27 M_\psi} \,.
\end{equation}
This expression requires only that the charm-quark mass is large compared to the strong-interaction scale. In the Coulombic limit, one may further use the result (\ref{fpsipsi}) for $f_\psi$. Note that the second term in (\ref{gammapsi}) constitutes a sizable fraction of about 20\% of the total rate, despite the electromagnetic origin of this contribution.

\section{A toy model}

Before we return to a discussion of $B\to X_s\,l^+l^-$ decays, we shall consider a toy model in which the hadronic part of the amplitude is exactly the current correlator discussed above, such that the role of resonances and quark-hadron duality is exhibited in a particularly transparent way. To this end, let us hypothetically assume the existence of two ``leptons'', $l_1$ with a large mass $m_1$ and $l_2$ with mass $m_2=0$, and the effective weak Hamiltonian
\begin{equation}\label{htoy}
   {\cal H}_{eff} 
   = \frac{G}{\sqrt{2}} \left[ 
   (\bar l_2l_1)_{V-A}\,(\bar cc)_{V-A} 
   - (\bar l_2l_1)_{V-A}\,(\bar tt)_{V-A} \right] .
\end{equation}
Since we are interested in the QCD dynamics of the decay, the flavour aspects of the model are unimportant for our discussion. All particles are assumed to have standard strong and electromagnetic interactions. Then ${\cal H}_{eff}$ gives rise to a loop-induced process $l_1\to l_2\,e^+e^-$ via charm- and top-quark penguin diagrams with a GIM-like cancellation between them. The corresponding decay amplitude reads 
\begin{equation}\label{al12ee}
   A(l_1\to l_2\,e^+e^-)
   = - \frac{G}{\sqrt{2}}\,e_c e^2\,\Pi(q^2)\,
   \bar l_2\gamma^\mu(1-\gamma_5)l_1\,\bar e\gamma_\mu e \,.
\end{equation}
Here $\Pi\equiv\Pi_c-\Pi_t$ is given as the difference between the charm and top contributions. We take $m_t > m_1$, and thus $ \mbox{Im}\,\Pi$ comes only from the charm sector. The correlator $\Pi(q^2)$ fulfills the dispersion relation (\ref{disprel}), where $\Pi(0)$ is fixed by our model and can be computed in perturbation theory. To leading order one finds
\begin{equation}\label{pi0}
   \Pi(0)\equiv \Pi_c(0)-\Pi_t(0)
   = \frac{N}{12\pi^2}\ln\frac{m^2_t}{m^2_c} \,.
\end{equation}
The form of (\ref{al12ee}) for the amplitude holds to lowest order in $G$ and $e^2$, but to all orders in the strong coupling. In fact, since the quark loops have no QCD interactions with the other, purely leptonic parts of the amplitude, the quantity $\Pi$ in (\ref{al12ee}) can be considered as the exact hadronic correlator in QCD. From (\ref{al12ee}) we obtain the differential decay rate (with $s=q^2/m_1^2$)
\begin{equation}\label{gaml12ee}
   \frac{d\Gamma(l_1\to l_2\,e^+e^-)}{ds}
   = \frac{G^2\alpha^2m^5_1}{27\pi}\,(1-s)^2\,(1+2s)\,
   \big| \Pi(q^2) \big|^2 .
\end{equation}

We shall now investigate to what extent the hadronic function $\Pi(q^2)$ in (\ref{gaml12ee}) may be approximated by a quark-level calculation. This is clearly the case for values of $q^2$ at which local duality is a reasonable approximation, that is if $q^2\gg 4m_c^2$ is well above the narrow-resonance region, or if $q^2<4m_c^2$ lies sufficiently far below the charmonium threshold
(we will make the latter statement more precise at the end of this section).
However, in a region of $q^2$ where prominent resonances exist, local duality is certainly badly violated. Then $|\Pi(q^2)|^2$ and also $\mbox{Im}\,\Pi(q^2)$ cannot be computed locally in the OPE approach. For $\mbox{Im}\,\Pi(q^2)$ the agreement of the full hadronic correlator with the quark-level calculation can be improved by integrating with a smooth weight function over a sufficiently large interval in $q^2$ (global duality). However, in (\ref{gaml12ee}) the correlator is {\em squared\/} before the $q^2$-integral can be performed to obtain the total rate for $l_1\to l_2\,e^+e^-$. As a consequence, the impact of sharp resonances is not averaged out even for the integrated rate. This is in fact our main point: The concept of global parton-hadron duality should not be expected to apply to averages of the quantity $|\Pi(q^2)|^2$.

To make this point clearer, we consider the contribution of a single $\psi$ resonance to $\Pi(q^2)$, omitting for the moment other resonances and non-resonant terms. The correlator then has the approximate form
\begin{equation}\label{pi1res}
   \Pi(q^2) = - \frac{f_\psi^2}{q^2-M_\psi^2+iM_\psi\Gamma_\psi} \,,
\end{equation}
where $f_\psi$, $M_\psi$, and $\Gamma_\psi$ denote the decay constant, mass, and total width of the resonance.\footnote{If $\Pi(q^2)$ is calculated within QCD, then strictly speaking  $\Gamma_\psi$ excludes the electromagnetic contribution in (\ref{gammapsi}) to the total decay width. In the following, it will be useful to consider $\Gamma_\psi$ as an adjustable parameter. For instance, we might consider a variant of the model, in which $\Pi(q^2)$ is the correlator of a current with different quark flavours, so that the QCD contribution to the width of the resonance is exactly zero. In this case $\Gamma_\psi$ is proportional to the arbitrary coupling constant of a hypothetical interaction by which the resonance decays to $e^+e^-$.}
It follows that
\begin{equation}\label{impi1res}
   \mbox{Im}\,\Pi(q^2)
   = \frac{f_\psi^2 M_\psi\Gamma_\psi}%
          {(q^2-M_\psi^2)^2+ M_\psi^2\Gamma_\psi^2} \,,
\end{equation}
while 
\begin{equation}\label{pi2im}
   \big| \Pi(q^2) \big|^2
   = \frac{f_\psi^4}{(q^2-M_\psi^2)^2+M_\psi^2\Gamma_\psi^2}
   = \frac{f_\psi^2}{M_\psi\Gamma_\psi}\,\mbox{Im}\,\Pi(q^2) \,.
\end{equation}
It is important to note that in the small-width limit $|\Pi(q^2)|^2$ is more singular than $\mbox{Im}\,\Pi(q^2)$.

Concentrating first on the absorptive part of the correlator, we note that it can deviate by a large factor from the partonic result. For instance, on resonance we have
\begin{equation}\label{impimax}
   \mbox{Im}\,\Pi(M^2_\psi)
   = \frac{f_\psi^2}{M_\psi\Gamma_\psi} 
   \ne \mbox{Im}\,\Pi_q(M^2_\psi) = {\cal O}(1) \,.
\end{equation}
At the $\psi$ resonance, for which $M_\psi=3.097\,{\rm GeV}$, 
$f_\psi=0.401\,{\rm GeV}$, and $\Gamma_\psi=93.2\cdot 10^{-6}$\,GeV, 
\begin{equation}\label{f2mgamnum}
   \frac{f^2_\psi}{M_\psi\Gamma_\psi}\approx 560 \,,
\end{equation}
corresponding to a huge violation of local duality. This behaviour has both a parametric and a numerical origin, as can be seen from (\ref{psiwidths}), 
which gives, using
$\alpha_s(M_\psi)=0.25$,
\begin{equation}\label{f2mgampar}
   \frac{f^2_\psi}{M_\psi\Gamma_\psi}
   \approx \frac{243}{40(\pi^2-9)\,\alpha^3_s(M_\psi)} 
   \approx 450 \,,
\end{equation}
in reasonable agreement with the experimental result (\ref{f2mgamnum}). In spite of this, the resonance contribution to the integral 
\begin{equation}\label{intimpi}
   \int_0^{m^2_1}\!dq^2\,\mbox{Im}\,\Pi(q^2)
   \approx \pi f_\psi^2\ll m^2_1
\end{equation} 
is parametrically small compared to the integral over the non-resonant contribution, which is of order $m_1^2$ when $m^2_1$ is large. This is consistent with global duality. In fact, the integral over the partonic expression for $\mbox{Im}\,\Pi(q^2)$ is dual to the integral over the hadronic expression including the contribution from the resonances. This property is used in the determination of the charm-quark mass from QCD sum rules \cite{Novikov:1976tn,Kuhn:2007vp}, as well as in the determination of $\alpha_s$ from $\tau$ decays \cite{Braaten:1991qm}, in which case it is applied to light-quark resonances.

We now contrast this to the situation for the square of the correlator. For the narrow $\psi$ resonance, using (\ref{pi2im}) and 
\begin{equation}\label{impidel}
   \mbox{Im}\,\Pi(q^2)\approx \pi f^2_\psi\,\delta(q^2-M^2_\psi) \,,
\end{equation}
we obtain for the resonance contribution to the integral
\begin{equation}\label{intpisq}
  \int_0^{m^2_1}\!dq^2\,|\Pi(q^2)|^2 
  \approx \pi f_\psi^2\times\frac{f_\psi^2}{M_\psi\Gamma_\psi} \,.
\end{equation} 
This should be compared with the non-resonant contribution, which remains of order $m_1^2$. The appearance of the factor $f_\psi^2/(M_\psi \Gamma_\psi)$ with the width in the denominator allows the resonance contribution to be arbitrarily large. Since expression (\ref{al12ee}) is exact in QCD, it is clear that global duality cannot possibly hold in this case and may be violated by an arbitrarily large amount (to the extent that we can adjust the resonance width at will). Substituting the resonance expression for $|\Pi(q^2)|^2$ into (\ref{gaml12ee}), we find
\begin{equation}\label{gaml12eeres}
   \Gamma_{\rm res}(l_1\to l_2\,e^+e^-)
   = \int_0^1\!ds\,\frac{d\Gamma_{\rm res}(l_1\to l_2\,e^+e^-)}{ds}
   = \Gamma(l_1\to l_2\psi)\,
   \frac{\Gamma(\psi\to e^+e^-)}{\Gamma_\psi} \,,
\end{equation}
where (with $r=M^2_\psi/m^2_1$)
\begin{equation}\label{gaml12psi}
   \Gamma(l_1\to l_2\psi)
   = \frac{G^2m^3_1 f^2_\psi}{16\pi}\,(1-r)^2\,(1+2r) \,.
\end{equation}
This reproduces the result for the rate of the decay chain $l_1\to l_2\psi\to l_2\,e^+e^-$ in the narrow-width approximation. 

The ratio $R_\psi$ of the resonance rate to the non-resonant partonic rate can now be estimated as follows. First, the partonic rate is obtained from (\ref{gaml12ee}) using the partonic expression for $\Pi(q^2)$. In the one-loop approximation we find
\begin{equation}\label{pipart}
   \frac{16\pi^2}{9}\Pi_q(q^2)
   = \frac{8}{9} \ln\frac{m_t}{m_c} + \frac{20}{27} + \frac{4}{9}\,x
   - \frac{2}{9}\,(2+x) \sqrt{|1-x|}\, \left\{
   \begin{array}{lc}
   2\arctan\frac{1}{\sqrt{x-1}} \,; &  x>1 \,, \\[3mm]
   \ln\frac{1+\sqrt{1-x}}{1-\sqrt{1-x}} - i\pi \,;~ & x<1 \,,
   \end{array}
   \right.
\end{equation} 
where $x=4m^2_c/q^2$. For typical values of the parameters (e.g.\ $m_c=1.4$\,GeV, $m_1=m_b=4.8$\,GeV, $m_t=167$\,GeV) the minimum of $|\Pi_q(q^2)|$ is located at $q^2=0$, while its maximum occurs at $q^2=4m^2_c$. At these values of $q^2$
\begin{equation}\label{pi04mc2}
   \big|\Pi_q(q^2)\big| 
   = \frac{1}{2\pi^2} \left( \ln\frac{m_t}{m_c} + c \right) ,
\end{equation}
with $c=0$ at $q^2=0$ and $c=4/3$ at $q^2=4m^2_c$. The total rate is well represented by treating $|\Pi_q(q^2)|$ as constant with an intermediate value of $c=1/2$. It can then be estimated to be 
\begin{equation}\label{gampart}
   \Gamma(l_1\to l_2\,e^+e^-)_{\rm SD}
   \approx \frac{G^2\alpha^2 m^5_1}{216\pi^5} 
   \left( \ln\frac{m_t}{m_c} + \frac{1}{2} \right)^2 .
\end{equation}
The resonant contribution is
\begin{equation}\label{gamres}
   \Gamma_{\rm res}(l_1\to l_2\psi\to l_2\,e^+e^-)
   = \frac{G^2\alpha^2 m^3_1 f^4_\psi}{27 M_\psi\Gamma_\psi}\,
   (1-r)^2\,(1+2r) \,.
\end{equation}
We thus find for their ratio
\begin{equation}\label{gamrespart}
   R_\psi = 
   \frac{\Gamma_{\rm res}(l_1\to l_2\psi\to l_2\,e^+e^-)}%
        {\Gamma(l_1\to l_2\,e^+e^-)_{\rm SD}}
   = \frac{8\pi^5(1-r)^2(1+2r)}{(\ln(m_t/m_c)+1/2)^2}
   \times \frac{f^2_\psi}{m^2_1}
   \times \frac{f^2_\psi}{M_\psi\Gamma_\psi}
   \approx 210 \,.
\end{equation}
In this example the first factor is about 55, $f^2_\psi/m^2_1=0.007$, and the last factor has been given in (\ref{f2mgamnum}). We observe that the moderate suppression of the channel $l_1\to l_2\psi$ by $55f^2_\psi/m^2_1\approx 0.38$ is overcompensated by the enhancement factor $f^2_\psi/(M_\psi\Gamma_\psi)$, whose origin we have discussed above in the context of (\ref{pi2im}). Since the latter factor is very large because of the narrow width of the $\psi$ resonance, we have $\Gamma_{\rm res}\gg\Gamma_{\rm SD}$. As anticipated, this result excludes the applicability of the quark-hadron duality hypothesis. 

\begin{figure}[t]
\begin{center}
\resizebox{13cm}{!}{\includegraphics{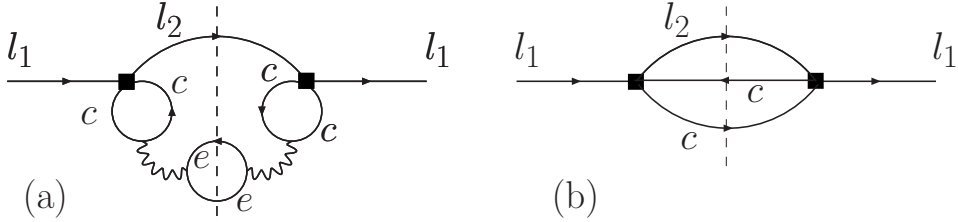}}
\end{center}
\caption{\label{fig:l1l2ee}
Cuts through the $l_1\to l_1$ forward-scattering diagrams that contribute (a) to the $l_1\to l_2\,e^+e^-$ decay rate and (b) to the inclusive hadronic decay rate for $l_1\to l_2+X$. See text for further explanation.}
\end{figure}

The deeper origin for the absence of global duality can be seen in the fact that there is no OPE for the total rate of $l_1\to l_2\,e^+e^-$ decay. Constructing the rate as the absorptive part of the $l_1\to l_1$ forward-scattering amplitude,
\begin{equation}\label{imhh}
   \langle l_1|\mbox{Im}\,i\int d^4x\,
   {\cal H}_{eff}(x)\,{\cal H}_{eff}(0)|l_1\rangle \,,
\end{equation}
we would need contributions with an $e^+e^-$ loop, coupled to charm- and top-quarks by photons, and we would have to require that the cut passes through this loop in order to obtain the rate for $l_1\to l_2\,e^+e^-$ (see the first graph in Figure~\ref{fig:l1l2ee}). This does not correspond to an OPE of (\ref{imhh}), which would imply a sum over all cuts, not just a restricted set. In this way $\Gamma(l_1\to l_2\,e^+e^-)$ is seen not to be a truly inclusive quantity, for which global duality would be expected to hold. Rather, the selection of a final state with an $e^+e^-$ pair represents a more ``exclusive'' choice, even for the integrated $l_1\to l_2\,e^+e^-$ rate, which leads to an integral over $|\Pi(q^2)|^2$. The situation encountered here is similar to that for the radiative decay $B\to X_s\gamma$. As emphasized in \cite{Lee:2006wn}, in this case contributions to the decay rate for which the photon is not part of the local operators in the effective weak Hamiltonian cannot be obtained from an OPE. They correspond to a subset of cuts analogous to those in Figure~\ref{fig:l1l2ee}.

On the other hand, a very different situation occurs for the inclusive hadronic decay $l_1\to l_2\,X$ (see the second graph in Figure~\ref{fig:l1l2ee}). In that case no restriction is placed on the cuts, and an OPE can be applied to (\ref{imhh}). Similarly to the rate of hadronic $\tau$ decay \cite{Braaten:1991qm}, the decay rate is given by a weighted integral over the imaginary parts of the vector-current ($\Pi$) and axial-vector current correlators ($\Pi^{T,L}_A$), 
\begin{equation}
   \Gamma(l_1\to l_2\,X)
   = \frac{G^2 m^5_1}{16\pi^2} \int_0^1\!ds\,(1-s)^2 
   \left[ (1+2s) \left( \mbox{Im}\,\Pi(q^2) + \mbox{Im}\,\Pi^T_A(q^2) 
   \right) + \mbox{Im}\,\Pi^L_A(q^2) \right] ,
\end{equation}
and global duality works in the same way as for the charm contribution to the $e^+e^-\to\mbox{hadrons}$ cross section.

Finally, we briefly return to the low-$q^2$ region in $l_1\to l_2\, e^+e^-$
decays mentioned after (\ref{gaml12ee}). Here a quark-level calculation is
justified if $q^2$ is sufficiently below the $\psi$ resonance.
We estimate $q^2_{\rm max}$, the maximum value of $q^2$, up to which a
quark-level calculation of $\Pi(q^2)$ can be trusted.
For $q^2$ close to zero the quark picture is reliable, and the
$\psi$-resonance contribution is only a small part in a hadronic
representation of $\Pi$. As $q^2$ gets close to $M^2_\psi$, 
the resonance contribution dominates the
correlator while the partonic result for $\Pi$ is too small. 
Therefore, as an estimate for $q^2_{\rm max}$ we use the 
point in $q^2$ where the one-loop partonic result equals the 
$\psi$-resonance contribution to $\Pi(q^2)-\Pi(0)$.
Using $[\Pi(q^2)-\Pi(0)]_\psi=(f^2_\psi/M^2_\psi)\, q^2/(M^2_\psi-q^2)$
and the partonic expression in (\ref{pipart}) close to threshold
$q^2=4m^2_c\approx M^2_\psi$, we obtain
$q^2_{\rm max}=M^2_\psi-3\pi^2 f^2_\psi/2\approx 7\,{\rm GeV^2}$.
Here we have considered the limit $f^2_\psi/M^2_\psi\ll 1$ in order to
obtain a simple analytic expression.
We then have $\Pi(q^2_{\rm max})-\Pi(0)=0.04$ for the one-loop result,
$0.05$ for the $\psi$-resonance contribution, and $0.06$ for the
full hadronic expression in the KS representation. Since $\Pi(q^2)-\Pi(0)$
is subdominant in comparison to $\Pi(0)$, given in (\ref{pi0}), 
the quark-level estimate of $\Pi(q^2_{\rm max})$ is still rather accurate.
A similar discussion applies to $B\to X_s\, l^+l^-$.

\section{\boldmath Charm resonances in $B\to X_s\,l^+l^-$ decays}

We now return to the discussion of $B\to X_s\,l^+l^-$, following the analysis of the toy model considered in the previous section. The branching fraction for the decay chain $B\to X_s\psi\to X_s\,l^+l^-$ is obtained by multiplying (\ref{bspsi}) with
\begin{equation}\label{psiee2}
   \mbox{B}(\psi\to l^+l^-)
   = \frac{\Gamma(\psi\to l^+l^-)}{\Gamma_\psi} \,,
    \qquad
   \Gamma(\psi\to l^+l^-)
   = \frac{16\pi\alpha^2 f^2_\psi}{27 M_\psi} \,,
\end{equation}
where the total rate $\Gamma_\psi$ is given in (\ref{gammapsi}). Replacing $a_2\to\kappa a_2\approx 0.26$ in (\ref{bspsi}) and using (\ref{bsee}), we then have for the ratio of the resonant and the partonic rate
\begin{equation}\label{bspsiee2}
   R_\psi \equiv
   \frac{\mbox{B}(B\to X_s\psi\to X_s\,l^+l^-)}%
        {\mbox{B}(B\to X_s\,l^+l^-)_{\rm SD}}
   = \frac{512\pi^5\kappa^2\,a^2_2\,(1-r)^2\,(1+2r)}%
          {9(\langle |C_9|^2\rangle + |C_{10}|^2)}
   \times \frac{f^2_\psi}{m^2_b} 
   \times \frac{f^2_\psi}{M_\psi\Gamma_\psi} \,.
\end{equation} 
The first factor is about 23 numerically. The first two factors give approximately 0.16. The large enhancement from $f^2_\psi/(M_\psi\Gamma_\psi)=560$ overcomes the suppression $\sim 0.16$, and we recover $R_\psi\approx 90$. This explains the size of $R_\psi$ already quoted in (\ref{bspsiee}). 

It is interesting to consider the heavy-quark limit $m_b$, $m_c\gg\Lambda_{\rm QCD}$ with $m_c/m_b$ fixed, where the $\psi$ resonance is asymptotically a Coulombic bound state. If we still assume $\alpha\ll\alpha_s$, then (\ref{fpsipsi}) and (\ref{psiwidths}) imply
\begin{equation}\label{rpsio1}
   R_\psi
   = \frac{512\pi^5\kappa^2\,a^2_2\,(1-r)^2\,(1+2r)}%
          {9(\langle |C_9|^2\rangle + |C_{10}|^2)}
   \times \frac{54}{5\pi(\pi^2-9)} 
   \left( \frac{\alpha_s(m_c v)}{\alpha_s(M_\psi)} \right)^3
   \times \frac{m_c^2}{m_b^2} \,.
\end{equation}
We see that formally $R_\psi={\cal O}(1)$ in the heavy-quark limit. However, the expression (\ref{rpsio1}) contains a large enhancement from numerical factors and from the running of $\alpha_s$ between the momentum scale of the charmonium bound state, entering $f_\psi$, and the scale $M_\psi$, which is relevant for $\Gamma_\psi$. Using $\alpha_s(m_c v)/\alpha_s(M_\psi)\approx 2$, the asymptotic formula (\ref{rpsio1}) gives $R_\psi\approx 60$ and thus reproduces the bulk of the resonance enhancement. In the heavy-quark limit a resummation of Coulomb ladders would automatically account for the large $\psi$-resonance contribution in the total rate of $B\to X_s\,l^+l^-$. Upon Dyson resummation of finite-width effects into the resonance propagator, expression (\ref{rpsio1}) is obtained.

The large charm-loop effect may be contrasted with the case of a light resonance $\rho$, where $\Gamma_\rho$, $M_\rho$, $f_\rho\sim\Lambda_{\rm QCD}$, and therefore the ratio $R_\rho$, defined in analogy to (\ref{bspsiee2}), is strongly suppressed. Estimating the first factor to be about 10--50
and using $f^2_\rho/(M_\rho\Gamma_\rho)$ $=$ $0.38$, one finds
\begin{equation}
   R_\rho\approx [10\mbox{--}50]\times \frac{f^2_\rho}{m^2_b}
   \times \frac{f^2_\rho}{M_\rho\Gamma_\rho}
   \approx [0.007\mbox{--}0.036] \,. 
\end{equation}
Even though we cannot expect quark-hadron duality to hold, in practice the $\rho$-resonance contribution is negligibly small. This scenario is relevant for up-quark penguins in $B\to X_{s,d}\,l^+l^-$ decays.

\section{\boldmath Penguins with charm in $B\to\pi\pi$ decays}

It is interesting to contrast the situation in the rare leptonic process $B\to X_s\,l^+l^-$ with the penguin amplitudes in exclusive hadronic decays such as $B\to\pi\pi$, where similar diagrams with intermediate charm quarks contribute \cite{Bander:1979px,Buras:1994pb,Ciuchini:1997hb}.

\begin{figure}[t]
\begin{center}
\resizebox{7cm}{!}{\includegraphics{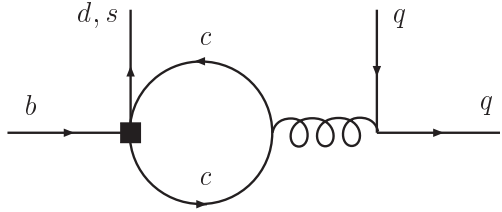}}
\end{center}
\caption{\label{fig:cpeng}
Penguin diagram with a charm-quark loop contributing to $B\to M_1 M_2$ decays. The curly line may either represent a gluon in the case of a QCD penguin graph, or a photon for an electromagnetic penguin.}
\end{figure}

Corrections of the penguin type as shown in Figure~\ref{fig:cpeng} enter the perturbatively calculable hard-scattering kernels for $B\to M_1 M_2$ decay amplitudes in QCD factorization. Such penguin contributions are consistently included in the factorization formula at next-to-leading order in perturbation theory, and at leading order in the heavy-quark limit \cite{BBNS1,BBNS2,BBNS3}. Nonperturbative strong interactions of the intermediate quarks and gluons do exist, but they are suppressed by powers of $\Lambda_{\rm QCD}/m_b$ with respect to the leading, factorizable amplitudes. In the following we shall discuss how the suppression comes about in the penguin contributions with a charm-quark or a light-quark loop. Of particular interest is the case of the charm penguin, where the validity of the usual factorization formula has been questioned in the literature \cite{Bauer:2004tj}.

In order to discuss the physics of penguin-type matrix elements with charm-loop contractions, it is instructive to consider first the case of electromagnetic penguins (Figure~\ref{fig:cpeng} with the gluon replaced by a photon). These are rather similar to the more prominent gluonic penguins, but the QCD dynamics is simpler in this case. The electromagnetic penguin matrix element contributes at ${\cal O}(\alpha)$ to the factorization coefficient $a^p_{10}$ ($p=u$, $c$) in the transition operator for $B\to\pi\pi$ decays, which contains the term \cite{BBNS3}
\begin{equation}\label{a10p}
   a_{10}^p\,\sum_q\,(\bar qb)_{V-A}
   \otimes \frac{3}{2}\,e_q\,(\bar dq)_{V-A} \,.
\end{equation}
The penguin contraction of the dominant current-current operators $Q_{1,2}^p$ gives a correction, included in $a^p_{10}$, which reads
\begin{equation}\label{dela10}
   \Delta a^p_{10}
   = \frac{C_1+ N C_2}{N}\,\frac{\alpha}{9\pi}
   \left[ - \frac{4}{3} \ln\frac{\mu}{m_b} + \frac{2}{3}
   - G_\pi(s_p) \right] ,
\end{equation}
where $s_p=m^2_p/m^2_b$ refers to the quark $p=u$, $c$ inside the loop, and
\begin{equation}\label{gpis}
   G_\pi(s) = \int_0^1\!dx\,G(s-i\epsilon,1-x)\,\phi_\pi(x)
\end{equation}
with
\begin{eqnarray}\label{gsx}
   G(s,\bar x) 
   &=& - 4\int_0^1\!du\,u(1-u) \ln\big[ s-u(1-u)\bar x \big] 
    \nonumber\\
   &=& - \frac{2}{3}\ln s + \frac{8\pi^2}{3}
    \left[ \Pi_q(\bar x m^2_b) - \Pi_q(0) \right]
\end{eqnarray}
is the penguin function, which involves a convolution of a perturbative hard-scattering kernel with the leading-twist pion distribution amplitude $\phi_\pi(x)$ \cite{BBNS1,BBNS2}. Eq.~(\ref{dela10}) represents the leading contribution of the electromagnetic penguin to the hard-scattering amplitude in QCD factorization, evaluated at lowest order in perturbation theory. The function $G_\pi(s_p)$ is an ${\cal O}(1)$ quantity, and $\Delta a^p_{10}={\cal O}(\alpha)$. Note the presence in the one-loop kernel $G(s,\bar x)$ of the vector-current correlator discussed in (\ref{piquark}). Indeed, the expression in square parentheses in (\ref{dela10}) can be rewritten as a term proportional to the convolution of the unsubtracted correlator $\Pi_q(\bar{x}m_b^2)$ with the pion distribution amplitude plus an additive correction of 2/3, which arises from the fact that in the NDR scheme with anticommuting $\gamma_5$, which we have adopted in \cite{BBNS3}, the four-dimensional Fierz identities are violated.

Consider now the effect of a $c\bar c$ resonance in the case of the charm-loop contribution. This corresponds to the diagram in Figure~\ref{fig:cpeng} (with photon exchange), but where the $c\bar c$ loop is replaced by a $\psi$ resonance propagator. Assuming factorization at the weak vertex, the contribution of the intermediate resonance to the decay amplitude may be evaluated in a straightforward way. For the equivalent of the function $G(s,\bar x)$ one finds, using the narrow-width approximation,
\begin{equation}\label{gpsi}
   G^\psi(\bar x)
   = - \frac{8\pi^2 f^2_\psi}{3 m^2_b}\,
   \frac{1}{\bar x - r_\psi + i\epsilon} \, ,
\end{equation}
where $r_\psi=M^2_\psi/m^2_b$. This may be integrated with the pion distribution amplitude $\phi_\pi(x)$ according to (\ref{gpis}). Employing the asymptotic form $\phi_\pi(x)=6x\bar x$ one obtains (for $0\leq r_\psi\leq 1$)
\begin{equation}\label{gpipsi}
   G^\psi_\pi
   = - \frac{8\pi^2 f^2_\psi}{m^2_b} \left[ 2 r_\psi\,(1-r_\psi)
   \left( \ln\frac{1-r_\psi}{r_\psi} - i\pi \right)
   + 1-2 r_\psi \right] .
\end{equation}
This contribution has to be considered as part of a hadronic representation of the hard-scattering amplitude $G_\pi$. It is suppressed in the heavy-quark limit. Consequently the existence of the resonance cannot alter substantially the quark-level result for $G_\pi$, neither parametrically nor 
numerically.\footnote{We estimate $G^\psi_\pi \approx -0.18+0.84 i$ for 
$m_b=4.8\,{\rm GeV}$, $f_\psi=0.401\,{\rm GeV}$. 
The partonic evaluation of the charm-loop 
contribution results in $G_\pi(s_c)= 2.32 + 1.19 i$ for 
$m_c=1.4\,{\rm GeV}$, and in $2.29 + 1.43 i$ for the smaller charm-mass 
value $m_c=1.25\,{\rm GeV}$. This should be compared to the result 
obtained from the hadronic representation of $\Pi(q^2)$  in the KS model 
\cite{Kruger:1996cv} giving $G^{KS}_\pi(s_c)= 1.86 + 2.02i$ and 
$2.01 + 2.02 i$ for the two charm-mass values, respectively.  
Thus, within uncertainties (perturbative, parametric, power corrections), 
the partonic and the hadronic (KS) results agree reasonably well. 
Note that the resonance contributions are more important for the 
imaginary part of $G_\pi(s_c)$ than for its real part. This is 
consistent with the behaviour of moment sum rules to determine the 
charm-quark mass, which are dominated by the charmonium resonances 
even for low moment order \cite{Kuhn:2007vp}. While the above numbers
are only estimates for the purpose of illustration, the analysis in 
\cite{Kuhn:2007vp} also indicates that global quark-hadron duality works
well for $\Pi(q^2)$ between $q^2=0$ and $q^2=m^2_b$.}
More specifically, in the Coulombic approximation the resonance contributes only at third order in $\alpha_s$, i.e.\ $G_\pi^\psi\sim\alpha^3_s\,(m_c/m_b)^2$, which follows from (\ref{fpsipsi}) and (\ref{gpipsi}). In the opposite limit, where the charm quark is treated as a light quark, we obtain instead $G_\pi^\psi\sim(\Lambda_{\rm QCD}/m_b)^2$. This demonstrates that the contribution of a light resonance is suppressed by two powers of $\Lambda_{\rm QCD}/m_b$. Note also that the imaginary part related to a light resonance has a suppression by four powers of $\Lambda_{\rm QCD}/m_b$ due to the additional endpoint suppression from the pion distribution amplitude in (\ref{gpipsi}).

An important difference between the cases $B\to X_s\,l^+l^-$ and $B\to\pi\pi$ results from the nature of the smearing procedure. In the first case the correlator is first squared and then integrated over phase-space, see (\ref{intpisq}). In the second case the contribution to the decay {\em amplitude\/} is given by an integral of a correlator with the pion distribution amplitude. This integral, for which duality holds, is then squared when one computes the decay rate. Note that the nature of the correlator itself is more complicated in the latter case. Instead of a correlation function of currents, the relevant correlator is the $\langle\pi\pi|\dots|B\rangle$ matrix element of the time-ordered product of a linear combination of the current-current operators $Q_{1,2}^c$ in the effective weak Hamiltonian with Standard Model interactions converting the $c\bar c$ pair into light quarks and gluons.

The assumption of factorization of the $c\bar c$ loop from the remaining quarks in the diagram is, of course, an idealization. However, gluon exchange between the charm loop and the $b$ and $s$ quarks does not invalidate our arguments. The most dangerous resonance contributions arise in the channel where the $c\bar c$ loop is in a colour-singlet state. Single-gluon exchange graphs are therefore less sensitive to such effects. In the case of the QCD penguin contribution the leading term in the QCD factorization approach contains the $c\bar c$ pair in a colour-octet state, and hence it is even less affected by resonance effects than the electroweak penguin contribution. When $m_c v^2 \sim \Lambda_{\rm QCD}$ is assumed, soft gluon exchange between the charm loop and the $b$ and $s$ quarks (hadronically, the remnant of the decaying $B$ meson) is expected to provide ${\cal O}(1)$ modifications in the resonance region, but it does not alter the power counting. Since the contribution from the resonance region is small, as shown above, the factorization result at leading power in the heavy-quark expansion remains unaffected.

The conclusion that the resonance contribution to the charm-penguin diagrams for exclusive hadronic $B$ decays into two light mesons is parametrically suppressed in the heavy-quark limit, irrespective of whether the charm quark is treated as being heavy or light, is in agreement with the general arguments on the suppression of the resonance or threshold region of $c\bar{c}$ loops given in \cite{Beneke:2004bn}. It disagrees with \cite{Bauer:2005wb} however, whose authors claim that QCD factorization fails for charm penguins even at leading power in the heavy-quark limit. We now explain why this claim is incorrect. The argument of \cite{Bauer:2005wb} is based on the estimate
\begin{equation}\label{eq:accbarscet}
   \frac{A_{c\bar c}}{A_{\rm LO}} \sim \alpha_s(2 m_c)\,
   f\bigg(\frac{2 m_c}{m_b}\bigg)\,v\, ,
\end{equation}
where $A_{c\bar c}$ is the non-perturbative contribution from the charm-quark penguins, $v$ is the small charm-quark velocity, and $f$ denotes some function of the ratio of heavy-quark masses. We find that there is an important factor missing on the right-hand side of (\ref{eq:accbarscet}). The matching described in the appendix of \cite{Bauer:2005wb} applies to the threshold region and is therefore done for a {\em fixed value\/} of $\bar u m_b^2$, the gluon virtuality in Figure~\ref{fig:cpeng}. The effective amplitude obtained in this way must then be integrated over the threshold region of size $\approx 4m_c^2 v^2$, which implies an integral of the form (with $r=4m_c^2/m_b^2$)
\begin{equation}\label{eq:ciprodint}
   \int_{r(1-v^2)}^{r(1+v^2)}\!du\,C_I^{\rm prod}(u)\,\dots \,,
\end{equation}
where $C_I^{\rm prod}$ is the coefficient function of the $c\bar c$ production operator defined in \cite{Bauer:2005wb}. Outside the threshold region the standard QCD factorization formulae apply and the factorization of charm loops is undisputed. Matching the amplitude gives $C_I^{\rm prod}\simeq 1$, and from the restricted integral in (\ref{eq:ciprodint}) we obtain a suppression factor of order $4m_c^2v^2/m_b^2$, by which the estimate in (\ref{eq:accbarscet}) should be multiplied to give
\begin{equation}\label{eq:accbarcorrect}
   \frac{A_{c\bar c}}{A_{\rm LO}} 
   \sim \alpha_s(2 m_c)\,f\bigg(\frac{2 m_c}{m_b}\bigg)\,v 
   \times \frac{4 m_c^2 v^2}{m_b^2} \,.
\end{equation}
This can easily be checked explicitly for the diagram in 
Figure~\ref{fig:cpeng}. 
One may also note from (\ref{gpipsi}) that the scaling of $\alpha G^\psi_\pi$ 
in the non-relativistic limit is of the same form as (\ref{eq:accbarcorrect}).
The reason for the omission in \cite{Bauer:2005wb} of the last factor in (\ref{eq:accbarcorrect}) can be traced to the fact that the 
authors incorrectly implement the restriction to the threshold region by 
assuming $C_I^{\rm prod}\simeq\delta{(\bar u-r)}$ 
and so miss the suppression from the small size of this region. 
Thus, contrary to a statement in their paper, the ``phase-space suppression factors'' discussed in \cite{Beneke:2004bn} are not included in (\ref{eq:accbarscet}). The analogy with inclusive quarkonium production mentioned in \cite{Bauer:2005wb} is misleading, since in this case factorization applies to the cross section, which contains the phase-space $\delta$-function in its definition, while here factorization is applied to an amplitude for fixed kinematics, which contains no $\delta$-function. The last factor in (\ref{eq:accbarcorrect}) suppresses non-perturbative effects in the charm-penguin contribution to the amplitude. Following \cite{Bauer:2005wb} and taking $m_c v^2\sim\Lambda_{\rm QCD}$, the additional factor is readily seen to give a power suppression, specifically $A_{c\bar{c}}/A_{\rm LO}\sim(\alpha_s/v)\,f(2m_c/m_b)\,(\Lambda_{\rm QCD}/m_b)^2$. If, on the other hand, we adopt the power counting $m_cv^2\gg\Lambda_{\rm QCD}$, then the non-relativistic scales can be treated in perturbation theory, and the uncalculable non-perturbative effects are again power suppressed \cite{Beneke:2004bn}.

We hope that this discussion clarifies that non-perturbative effects in the threshold region in charm-penguin diagrams are power suppressed in the heavy-quark limit and hence do not spoil QCD factorization.

\section{Conclusions}

We have explained the origin of the well-known fact that the integrated $B\to X_s\,l^+l^-$ branching fraction is dominated by resonance contributions from narrow charmonium states, such as $B\to X_s\psi\to X_s\,l^+l^-$, which exceed the non-resonant charm-loop contribution by two orders of magnitude. Our arguments, which are based on well-established theoretical tools, show that quark-hadron duality cannot be expected to hold for these decays. On the other hand, we have shown that corresponding resonance effects lead to highly suppressed contributions to charm-penguin amplitudes in two-body hadronic $B$ decays of the type $B\to\pi\pi$, which do not invalidate the standard picture of QCD factorization. 

The contributions of charmonium states to the different observables studied in this note are described in terms of resonance contributions to a correlation function $\Pi(q^2)$. The question of when large deviations from quark-hadron duality arise is tightly linked to how this correlator enters the formulae for the various decay rates. In fully inclusive quantities such as the $c\bar c$ contribution to the $e^+e^-\to\mbox{hadrons}$ cross section, the imaginary part of a current-current correlator is integrated over phase space, and duality holds in the sense of \cite{Poggio:1975af}. A similar situation is realized in inclusive semileptonic or hadronic $B$-meson decays, not discussed here. The inclusive decay rates are given in terms of forward $B$-meson matrix elements of the imaginary parts of correlators of two effective weak Hamiltonians \cite{Bigi:1993fe,Manohar:1993qn}.

In more complicated cases such as the penguin contributions to exclusive hadronic decays $B\to M_1 M_2$ in the heavy-quark limit, the charm-penguin contributions to the decay amplitudes are given in terms of hadronic matrix elements of time-ordered products of current-current operators from the effective weak Hamiltonian with Standard Model Lagrangian insertions, convoluted with the pion distribution amplitude. Even though the QCD dynamics is considerably more complicated in this case, the resonance physics is similar, and duality still holds in a global sense, that the integrated (smeared) correlation functions calculated in a partonic picture approximate the true hadronic correlation functions in the heavy-quark limit. 

The charm-resonance contributions to $B\to X_s\,l^+l^-$ decays, on the other hand, are expressed in terms of a phase-space integral over the absolute {\em square\/} of a correlator. We have shown that due to the fact that the widths of the lowest-lying charmonium states are very narrow, this integral violates the duality hypothesis by an amount that becomes arbitrarily large in the limit $\Gamma_\psi\to 0$. Our main conclusions are derived with the example of a toy model, which keeps the essential features of the resonance physics while simplifying other aspects of the decay processes.

Our discussion in this note has concentrated on the inclusive leptonic process $B\to X_s\,l^+l^-$, for which duality violations in the charm resonance region are particularly pronounced and experimentally well studied. An analogous discussion applies to the related radiative process $B\to X_s\gamma$.

\subsubsection*{Acknowledgements}

We thank Tobias Hurth, Gino Isidori and Miko{\l}aj Misiak for discussions
on the charm-loop contribution in $B\to X_s\,l^+l^-$, and Tobias Hurth for 
his persistent encouragement to write this paper. M.B.\ and G.B.\ thank the 
CERN Theory Group and C.T.S.\ thanks the KEK Theory Group for their 
hospitality while this work was completed. This work is supported in part by 
the DFG Sonder\-forschungs\-be\-reich/Trans\-regio~9 ``Computergest\"utzte 
Theoretische Teilchenphysik'', the DFG cluster of excellence
``Origin and Structure of the Universe'', the excellence research center 
``Elementarkr\"afte und ma\-the\-ma\-tische Grundlagen'' at Mainz University, 
the STFC Grant ST/G000557/1, and the EU contract MRTN-CT-2006-035482 
(Flavianet).


\begin{thebibliography}{99}

\bibitem{Poggio:1975af}
  E.~C.~Poggio, H.~R.~Quinn and S.~Weinberg,
  %``Smearing The Quark Model,''
  Phys.\ Rev.\  D {\bf 13} (1976) 1958.
  %%CITATION = PHRVA,D13,1958;%%

\bibitem{BBNS1}
  M.~Beneke, G.~Buchalla, M.~Neubert and C.~T.~Sachrajda,
  %``{QCD} factorization for B --> pi pi decays:
  %Strong phases and CP  violation in the heavy quark limit,''
  Phys.\ Rev.\ Lett.\ {\bf 83} (1999) 1914
  [hep-ph/9905312].
  %%CITATION = HEP-PH 9905312;%%

\bibitem{BBNS2}
  M.~Beneke, G.~Buchalla, M.~Neubert and C.~T.~Sachrajda,
  %``QCD factorization for exclusive, non-leptonic B meson decays:
  %General  arguments and the case of heavy-light final states,''
  Nucl.\ Phys.\  {\bf B591} (2000) 313
  [hep-ph/0006124].
  %%CITATION = HEP-PH 0006124;%%

\bibitem{BBNS3}
  M.~Beneke, G.~Buchalla, M.~Neubert and C.~T.~Sachrajda,
  %``QCD factorization in B $\to$ pi K, pi pi decays and extraction of
  %Wolfenstein parameters,''
  Nucl.\ Phys.\ B {\bf 606} (2001) 245
  [hep-ph/0104110].
  %%CITATION = HEP-PH 0104110;%%

\bibitem{BBL}
  G.~Buchalla, A.~J.~Buras and M.~E.~Lautenbacher,
  %``Weak Decays Beyond Leading Logarithms,''
  Rev.\ Mod.\ Phys.\  {\bf 68} (1996) 1125 
  [hep-ph/9512380].
  %%CITATION = HEP-PH 9512380;%%

\bibitem{Amsler:2008zzb}
  C.~Amsler {\it et al.}  [Particle Data Group],
  %``Review of particle physics,''
  Phys.\ Lett.\  B {\bf 667} (2008) 1.
  %%CITATION = PHLTA,B667,1;%%

\bibitem{Ko:1995iv}
  P.~Ko, J.~Lee and H.~S.~Song,
  %``Inclusive $S-$wave charmonium productions in $B$ decays,''
  Phys.\ Rev.\  D {\bf 53} (1996) 1409
  [hep-ph/9510202].
  %%CITATION = PHRVA,D53,1409;%%

\bibitem{Beneke:1998ks}
  M.~Beneke, F.~Maltoni and I.~Z.~Rothstein,
  %``{QCD} analysis of inclusive B decay into charmonium,''
  Phys.\ Rev.\  D {\bf 59} (1999) 054003
  [hep-ph/9808360].
  %%CITATION = PHRVA,D59,054003;%%

\bibitem{Kruger:1996cv}
  F.~Kr\"uger and L.~M.~Sehgal,
  %``Lepton Polarization in the Decays $B\to X_s\mu~+\mu~-$ and $B\to
  %X_s\tau~+\tau~-$,''
  Phys.\ Lett.\  B {\bf 380} (1996) 199
  [hep-ph/9603237].
  %%CITATION = PHLTA,B380,199;%%

\bibitem{Chibisov:1996wf}
  B.~Chibisov, R.~D.~Dikeman, M.~A.~Shifman and N.~Uraltsev,
  %``Operator product expansion, heavy quarks, QCD duality and its
  %violations,''
  Int.\ J.\ Mod.\ Phys.\  A {\bf 12} (1997) 2075
  [hep-ph/9605465].
  %%CITATION = IMPAE,A12,2075;%%

\bibitem{BSZ}
  B.~Blok, M.~A.~Shifman and D.~X.~Zhang,
  %``An illustrative example of how quark-hadron duality might work,''
  Phys.\ Rev.\ D {\bf 57} (1998) 2691
  [Erratum-ibid.\ D {\bf 59} (1999) 019901] 
  [hep-ph/9709333].
  %%CITATION = HEP-PH 9709333;%%

\bibitem{Sommerfeld}
  A.~Sommerfeld, Annalen der Physik {\bf 403} (1931) 257.

\bibitem{Beneke:1999zr}
  M.~Beneke, {\it Perturbative heavy quark-antiquark systems}, 
  [hep-ph/9911490], in: Proceedings of the 8th International Symposium on 
  Heavy Flavor Physics (Heavy Flavours 8), Southampton, UK (July 1999).
  %%CITATION = HEP-PH/9911490;%%

\bibitem{Voloshin}
  M.~B.~Voloshin, Sov.\ J.\ Nucl.\ Phys.\ {\bf 36} (1982) 143 
  [Yad.\ Fiz.\ {\bf 36} (1982) 247]. 

\bibitem{Novikov:1976tn}
  V.~A.~Novikov, L.~B.~Okun, M.~A.~Shifman, A.~I.~Vainshtein, 
  M.~B.~Voloshin and V.~I.~Zakharov,
  %``Sum Rules For Charmonium And Charmed Mesons Decay Rates In Quantum
  %Chromodynamics,''
  Phys.\ Rev.\ Lett.\  {\bf 38} (1977) 626
  [Erratum-ibid.\  {\bf 38} (1977) 791].
  %%CITATION = PRLTA,38,626;%%

\bibitem{Kuhn:2007vp}
  J.~H.~K\"uhn, M.~Steinhauser and C.~Sturm,
  %``Heavy quark masses from sum rules in four-loop approximation,''
  Nucl.\ Phys.\  B {\bf 778} (2007) 192
  [hep-ph/0702103].
  %%CITATION = NUPHA,B778,192;%%

\bibitem{Braaten:1991qm}
  E.~Braaten, S.~Narison and A.~Pich,
  %``QCD analysis of the tau hadronic width,''
  Nucl.\ Phys.\  B {\bf 373} (1992) 581.
  %%CITATION = NUPHA,B373,581;%%

\bibitem{Lee:2006wn}
  S.~J.~Lee, M.~Neubert and G.~Paz,
  %``Enhanced non-local power corrections to the B --> X/s+ gamma decay rate,''
  Phys.\ Rev.\  D {\bf 75} (2007) 114005
  [hep-ph/0609224].
  %%CITATION = PHRVA,D75,114005;%%

\bibitem{Bander:1979px}
  M.~Bander, D.~Silverman and A.~Soni,
  %``CP Noninvariance In The Decays Of Heavy Charged Quark Systems,''
  Phys.\ Rev.\ Lett.\  {\bf 43} (1979) 242.
  %%CITATION = PRLTA,43,242;%%

\bibitem{Buras:1994pb}
  A.~J.~Buras and R.~Fleischer,
  %``Limitations in measuring the angle Beta by using SU(3) relations for B
  %meson decay amplitudes,''
  Phys.\ Lett.\  B {\bf 341} (1995) 379
  [hep-ph/9409244].
  %%CITATION = PHLTA,B341,379;%%

\bibitem{Ciuchini:1997hb}
  M.~Ciuchini, E.~Franco, G.~Martinelli and L.~Silvestrini,
  %``Charming penguins in B decays,''
  Nucl.\ Phys.\  B {\bf 501} (1997) 271
  [hep-ph/9703353].
  %%CITATION = NUPHA,B501,271;%%

\bibitem{Bauer:2004tj}
  C.~W.~Bauer, D.~Pirjol, I.~Z.~Rothstein and I.~W.~Stewart,
  %``B --> M(1) M(2): Factorization, charming penguins, strong phases, and
  %polarization,''
  Phys.\ Rev.\  D {\bf 70} (2004) 054015
  [hep-ph/0401188].
  %%CITATION = PHRVA,D70,054015;%%

\bibitem{Beneke:2004bn}
  M.~Beneke, G.~Buchalla, M.~Neubert and C.~T.~Sachrajda,
  %``Comment on 'B --> M(1) M(2): Factorization, charming penguins, strong
  %phases, and polarization',''
  Phys.\ Rev.\  D {\bf 72} (2005) 098501
  [hep-ph/0411171].
  %%CITATION = PHRVA,D72,098501;%%

\bibitem{Bauer:2005wb}
  C.~W.~Bauer, D.~Pirjol, I.~Z.~Rothstein and I.~W.~Stewart,
  %``On differences between SCET and QCDF for B --> pi pi decays,''
  Phys.\ Rev.\  D {\bf 72} (2005) 098502
  [hep-ph/0502094].
  %%CITATION = PHRVA,D72,098502;%%

\bibitem{Bigi:1993fe}
  I.~I.~Y.~Bigi, M.~A.~Shifman, N.~G.~Uraltsev and A.~I.~Vainshtein,
  %``QCD predictions for lepton spectra in inclusive heavy flavor decays,''
  Phys.\ Rev.\ Lett.\  {\bf 71}, 496 (1993)
  [hep-ph/9304225].
  %%CITATION = PRLTA,71,496;%%

\bibitem{Manohar:1993qn}
  A.~V.~Manohar and M.~B.~Wise,
  %``Inclusive semileptonic B and polarized Lambda(b) decays from QCD,''
  Phys.\ Rev.\  D {\bf 49} (1994) 1310
  [hep-ph/9308246].
  %%CITATION = PHRVA,D49,1310;%%

\end{thebibliography}
\end{document}